\documentclass[12pt]{article}
\usepackage{graphics}
\begin{document}

\title{Magnetic field effects on the optimal fidelity of standard teleportation
via the two qubits Heisenberg XX chain in thermal equilibrium}
\author{Shang-Bin Li and Jing-Bo Xu\\
Zhejiang Institute of Modern Physics and Department of Physics,\\
Zhejiang University, Hangzhou 310027, People's Republic of
China\thanks{Mailing address}}
\date{}

\maketitle

\begin{abstract}
{\normalsize We study the two qubits Heisenberg XX chain with
magnetic impurity in the presence of the external magnetic field
and calculate the optimal fidelity of standard teleportation via
the thermal equilibrium state. It is found that the combined
influence of magnetic impurity and external magnetic field can
increase the critical temperatures of entanglement and quantum
teleportation without limit. The relation between two kinds of
critical temperatures is revealed.\\

}

\end{abstract}
\newpage

\hspace*{8mm}In recent years, considerable attention has been
devoted to the quantum entanglement [1-4] and teleportation
[5,6,7] in Heisenberg spin chains, which is closely associated
with the study of many practical systems and can be used to
realize universal gate operations in solid state quantum
computation [8]. The thermal entanglement of spins in a Heisenberg
chain with an external magnetic field in thermal equilibrium with
a nonzero temperature has been studied [1]. It is shown that the
entanglement of a two qubit isotropic Heisenberg system decreases
with increasing the temperature $T$ and vanishes beyond a critical
value $T_c$ [1]. The teleportation via the thermal states of the
two-qubit Heisenberg XX chain has also been investigated [7].
There is a critical temperature above which the thermal state of
two-qubit Heisenberg XX chain is useless for quantum standard
teleportation. In this Letter, we study the two qubits Heisenberg
XX chain with magnetic impurity in the presence of the external
magnetic field and calculate the entanglement and the optimal
fidelity of standard teleportation via the thermal equilibrium
state. It is found that the combined influence of magnetic
impurity and external magnetic field can increase the critical
temperatures of entanglement and standard teleportation. We
construct the envelope of the critical temperature of
teleportation fidelity, which is just as the same
as the critical temperature of entanglement.\\
\hspace*{8mm}The Hamiltonian for the two qubit Heisenberg XX chain
in an external magnetic field $B$ (along the $z$ axis) with a
magnetic impurity $B_1$ in the first qubit can be expressed as [9]
$$
H=(B+B_1)S^{1}_z+BS^{2}_z+J(S^{1}_{+}S^{2}_{-}+S^{1}_{-}S^{2}_{+}),
\eqno{(1)}
$$
where $S^{i}_{\pm}=S^{i}_x\pm{i}S^{i}_y$,
$S^{i}_{\alpha}=\sigma^{i}_{\alpha}/2$ ($\alpha=x,y,z$ and
$i=1,2$) denotes the local spin $\frac{1}{2}$ operator of the
$i$th qubit and $\sigma^{i}_{\alpha}$ are the Pauli operators. The
chain is said to be antiferromagnetic for the coupling constant
$J>0$ and ferromagnetic for $J<0$. The eigenvectors and
eigenvalues of the Hamiltonian $H$ are
$$
|\Psi_1\rangle=|00\rangle,~~~E_1=-B-\frac{1}{2}B_1,
$$
$$
|\Psi_2\rangle=|11\rangle,~~~E_2=B+\frac{1}{2}B_1,
$$
$$
|\Psi_3\rangle=\frac{J}{\sqrt{2\eta^2-B_1\eta}}|10\rangle+\frac{\eta-B_1/2}{\sqrt{2\eta^2-B_1\eta}}|01\rangle,~~~E_3=\eta,
$$
$$
|\Psi_4\rangle=\frac{J}{\sqrt{2\eta^2+B_1\eta}}|10\rangle-\frac{\eta+B_1/2}{\sqrt{2\eta^2+B_1\eta}}|01\rangle,~~~E_4=-\eta,
\eqno{(2)}
$$
where $\eta=\sqrt{J^2+B^2_1/4}$~. From Eq.(2), we can see that the
ground state of $H$ is just the Bell singlet state if $B_1=0$ and
$B^2<J^2$. For the system in thermal equilibrium at temperature
$T$, the density operator can be derived as
$$
\rho_{AB}=\frac{1}{Z}[u|11\rangle\langle11|+v|00\rangle\langle00|
+w_1|10\rangle\langle10|+w_2|01\rangle\langle01|+y(|10\rangle\langle01|+|01\rangle\langle10)],
\eqno{(3)}
$$
where
$$
u=\exp[-\beta(B+\frac{1}{2}B_1)],~~~v=\exp[\beta(B+\frac{1}{2}B_1)],
$$
$$
w_1=J^2[\frac{\exp(-\eta\beta)}{2\eta^2-\eta{B_1}}+\frac{\exp(\eta\beta)}{2\eta^2+\eta{B_1}}],
$$
$$
w_2=\frac{\exp(-\eta\beta)(\eta-B_1/2)^2}{2\eta^2-\eta{B_1}}+\frac{\exp(\eta\beta)(\eta+B_1/2)^2}{2\eta^2+\eta{B_1}},
$$
$$
y=-\frac{J}{\eta}\sinh(\eta\beta),~~~Z=2\cosh[(B+\frac{1}{2}B_1)\beta]+2\cosh(\eta\beta),
\eqno{(4)}
$$
and $\beta=\frac{1}{k_BT}$ with $k_B$ the Boltzmann's constant.
Next, we briefly investigate the entanglement characterized by
concurrence [10] of the two-qubit Heisenberg XX chain with
magnetic impurity in an external magnetic field in the thermal
equilibrium. The concurrence $C$ of density operator $\rho_{AB}$
can be calculated,
$$
C=2\max[0,(|y|-1)/Z], \eqno{(5)}
$$
where $|x|$ gives the absolute value of $x$. From Eq.(5), we can
see that the thermal entanglement is invariant under the
substitution $B\rightarrow{-B}$ and $B_1\rightarrow{-B_1}$ or
$J\rightarrow{-J}$. The latter indicates that the entanglement is
the same for the antiferromagnetic and ferromagnetic cases. For
$T=0$, at certain critical value $B^{\pm}_c$ of the magnetic field
$B$, the entanglement becomes a nonanalytical function of the
magnetic field and a quantum phase transition occurs [1,11]. It is
easy to see that the critical magnetic field $B^{\pm}_c$ is
dependent of the magnetic impurity $B_1$ and can be expressed as
$B^{\pm}_c=\eta\pm\frac{1}{2}B_1$.\\
\begin{figure}
\centering
\includegraphics{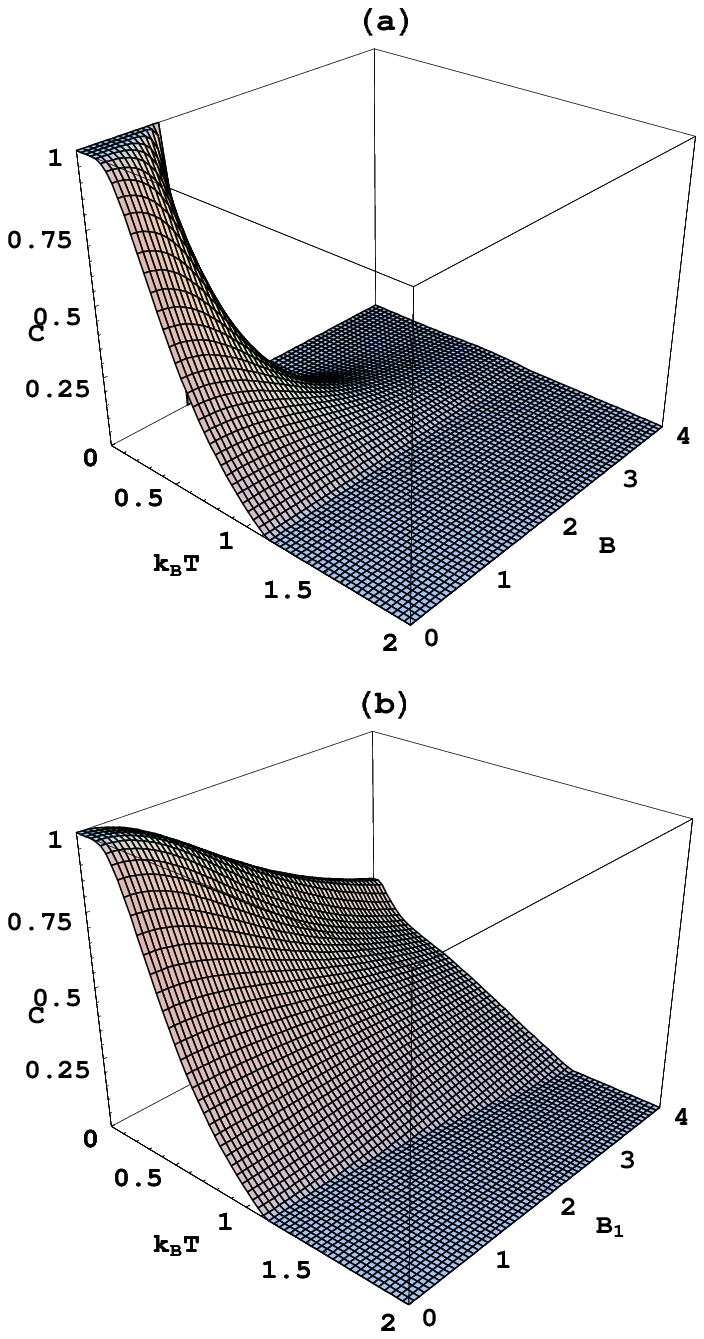}
\caption{(a) The concurrence $C$ is plotted as a function of
$k_BT$ and $B$ with $J=1$ and $B_1=0$; (b) the concurrence $C$ is
plotted as a function of $k_BT$ and $B_1$ with $J=1$ and $B=0$.
\label{Fig1}}
\end{figure}
In Fig.1(a), the concurrence $C$ is plotted as a function of
$k_BT$ and $B$ with $J=1$ and $B_1=0$. In Fig.1(b), the
concurrence $C$ is plotted as a function of $k_BT$ and $B_1$ with
$J=1$ and $B=0$. From Fig.1(a), we can see that there exists a
critical temperature $k_BT_c\approx1.13459J$, beyond which the
thermal entanglement is zero. However, in Fig.1(b), it is shown
that the critical temperature is increased due to the presence of
the magnetic impurity. In fact, the critical temperature can be
explicitly expressed as
$$
k_BT_c=\frac{\eta}{\ln(\eta+\sqrt{J^2+\eta^2})-\ln{J}}. \eqno{(6)}
$$
In the following, we consider the standard teleportation protocol
by making use of the above two-qubit thermal state $\rho_{AB}$ as
resource. Horodecki et al. have defined a optimal fidelity
$f(\rho)$ of the standard teleportation scheme [12] which
quantifying the quality of the teleportation that can be achieved
with the given state $\rho$. The optimal fidelity of standard
teleportation is related with the maximal singlet fraction via the
equation $f(\rho)=\frac{2F(\rho)+1}{3}$, in which the maximal
singlet fraction $F(\rho)$ is defined as the maximal overlap of
the state $\rho$ with a maximally entangled (ME) state [12]
$$
F(\rho)=\max_{|\psi\rangle={\mathrm{ME}}}\langle\psi|\rho|\psi\rangle.
\eqno{(7)}
$$
An explicit value for the maximal singlet fraction has been
derived [13]. If one considers the real $3\times3$ matrix
$\tilde{R}={\mathrm{Tr}}(\rho\sigma_i\otimes\sigma_j)$ with
$\{\sigma_i,~i=1,2,3\}$ the Pauli matrices, then
$$
F(\rho)=\frac{1+\lambda_1+\lambda_2-{\mathrm{sgn}}[{\mathrm{det}}(\tilde{R})]\lambda_3}{4}
\eqno{(8)}
$$
with $\{\lambda_i\}$ the ordered singular values of $\tilde{R}$
and ${\mathrm{sgn}}[{\mathrm{det}}(\tilde{R})]$ the sign of the
determinant of $\tilde{R}$. According to Eq.(8), the maximal
singlet fraction $F(\rho_{AB})$ of the two-qubit thermal state
$\rho_{AB}$ is found to be,
$$
F(\rho_{AB})=\max\{\frac{\cosh[\beta(B+B_1/2)]}{Z},\frac{\eta\cosh\eta\beta+|J|\sinh\eta\beta}{\eta{Z}}\}
\eqno{(9)}
$$
\begin{figure}
\centering
\includegraphics{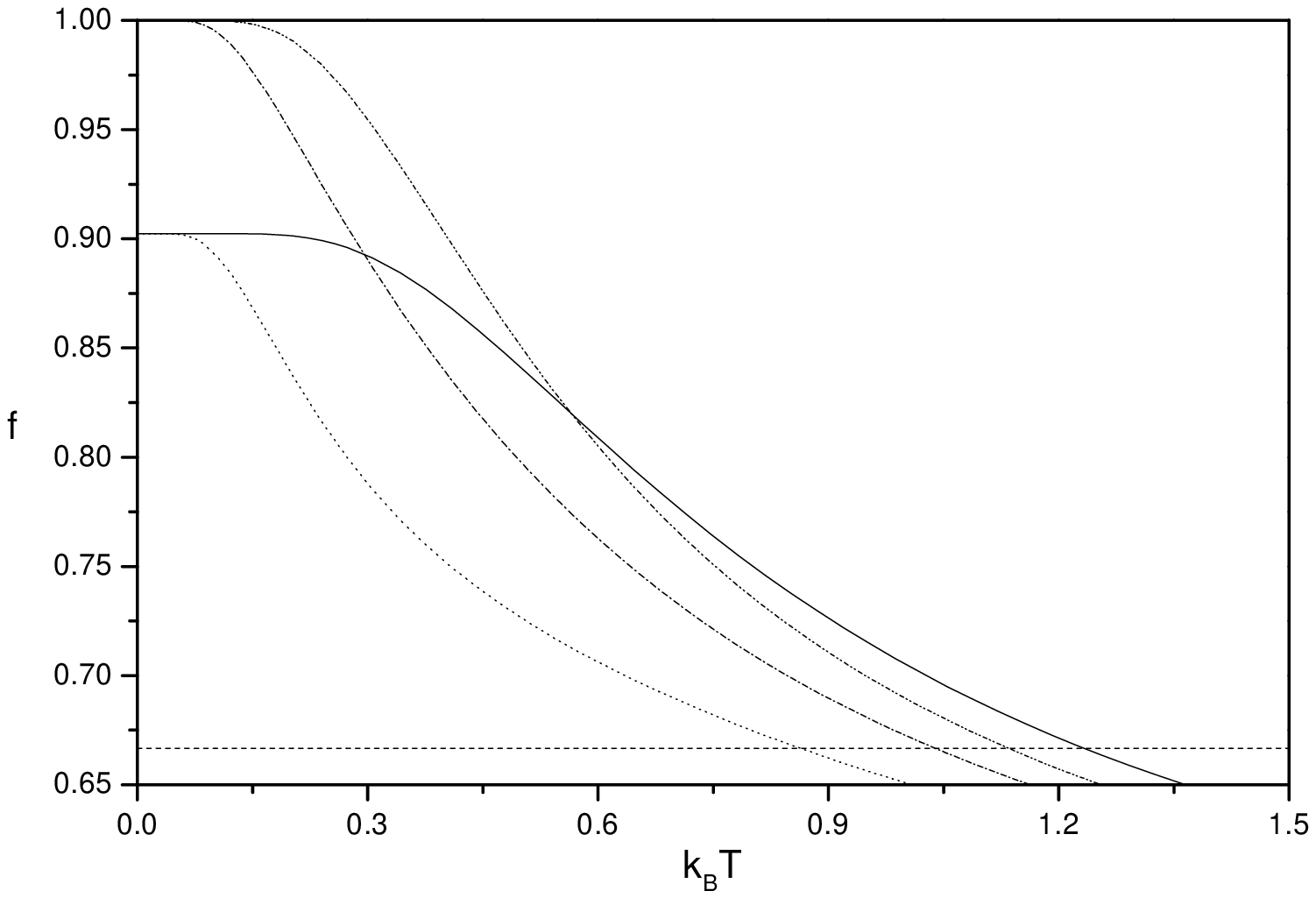}
\caption{The optimal fidelity $f$ of the thermal state $\rho_{AB}$
is plotted as a function of $k_BT$ with $J=1$ for four different
cases: (Solid Line) $B=-1$ and $B_1=2$; (Dot Line) $B=0$ and
$B_1=2$; (Dash Dot Line) $B=-0.5$ and $B_1=0$; (Dash Dot Dot Line)
$B=0$ and $B_1=0$ .\label{Fig2}}
\end{figure}
In Fig.2, the optimal fidelity $f$ of the thermal state
$\rho_{AB}$ is plotted as a function of $k_BT$ with $J=1$ for four
different cases. In the case of $B=0$ and $B_1=0$ (Dash Dot Dot
line in Fig.2), we can see that there exists a temperature
$T^{(f)}_c(0,0)$, beyond which the optimal fidelity is smaller
than the classical limit $2/3$. This critical temperature
$T^{(f)}_c(0,0)$ is given by $T^{(f)}_c(0,0)\simeq1.13459J/k_B$.
The critical temperature decreases with the external magnetic
field (See dash dot line in Fig.2) or the magnetic impurity (See
dot line in Fig.2). However, the combined influence of the
external magnetic field and the magnetic impurity can increase the
critical temperature $T^{(f)}_c$ above $1.13459J/k_B$ (See solid
line in Fig.2). It is easy to verify that the critical temperature
$T^{(f)}_c$ is the same as $T_c$
when $B_1=-2B$.\\
\begin{figure}
\centering
\includegraphics{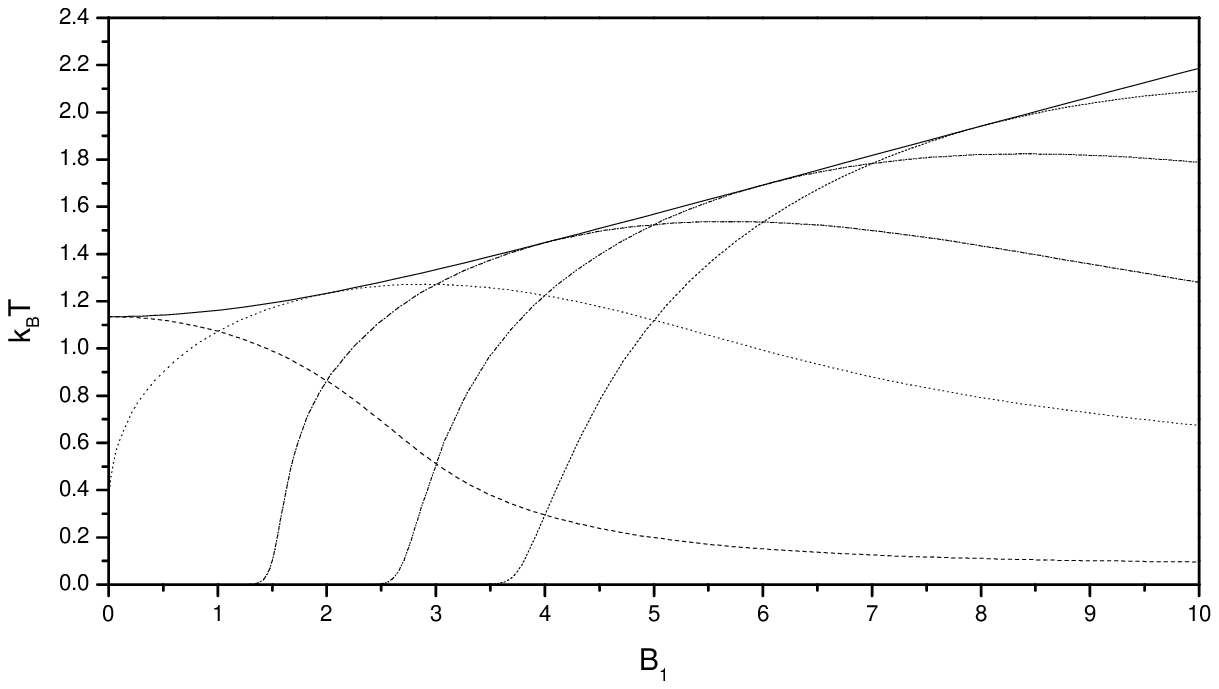}
\caption{ The critical temperature $k_BT_c$ (Solid Line) of
entanglement is plotted as a function of the magnetic impurity
$B_1$ with $J=1$. The critical temperature $k_BT^{(f)}_c$ of the
teleportation fidelity is plotted as a function of the magnetic
impurity $B_1$ with $J=1$ for five different values of external
magnetic field: (Dash Line) $B=0$; (Dot Line) $B=-1$; (Dash Dot
Line) $B=-2$; (Dash Dot Dot Line) $B=-3$; (Short Dash Line)
$B=-4$. \label{Fig3}}
\end{figure}\\
In Fig.3, the critical temperature $T_c$ of entanglement is
plotted as a function of the magnetic impurity $B_1$ with $J=1$
and the critical temperature $T^{(f)}_c$ of the teleportation
fidelity is plotted as a function of the magnetic impurity $B_1$
with $J=1$ for five different values of external magnetic field.
It is shown that the critical value $T_c$ increases with the
magnetic impurity $B_1$ and is independent of $B$. From Fig.3, it
is easy to see that the curve of $T_c$ constructs a envelope of
the curves $\{T^{f}_c(J,B_1,B)\}_{B}$. To prove it, we recall the
classical theory of envelopes [14]: Suppose one has a set of
functions $g(x;\alpha)$ parametrized by $\alpha$. Then the
envelope of $\{g(x;\alpha)\}_{\alpha}$ can be obtained by solving
$\partial{g}(x;\alpha)/\partial\alpha=0$ and substituting the
solution $\alpha(x)$ into the original function, i.e.,
$g(x;\alpha(x))$. In our case, $g=T^{(f)}_c$ and $\alpha=B$. Thus
we reach a condition
$$
\frac{\partial{T^{(f)}_c}(J,B_1,B)}{\partial{B}}=0.
\eqno{(10)}
$$
From Eq.(9), we know that $T^{(f)}_c$ satisfies the following
equation
$$
\sinh(\eta/k_BT^{(f)}_c)=\frac{\eta}{J}\cosh[(\frac{1}{2}B_1+B)/k_BT^{(f)}_c].
\eqno{(11)}
$$
Combining Eq.(10) and Eq.(11), we can derive that
$$
B=-\frac{1}{2}B_1
\eqno{(12)}
$$
is the unique solution. Substituting Eq.(12) into Eq.(11), it is
easy to verify that $T^{(f)}_c(J,B_1,-\frac{1}{2}B_1)$ is equal to
$T_c(J,B_1)$.

In conclusion, we studied the two-qubit Heisenberg XX chain with
magnetic impurity. Firstly, we investigated the entanglement in
the two-qubit Heisenberg XX chain with magnetic impurity, and it
is shown that the critical temperature $T_c$ of entanglement,
which is independent of the external magnetic field, can be
improved with the increasing intensity of magnetic impurity. Then,
We show that the combined influence of magnetic impurity and the
external magnetic field can improve the critical temperature
(beyond which the optimal fidelity is smaller than the classical
limit $2/3$) of thermal state of the two-qubit Heisenberg XX chain
as a useful resource for standard quantum teleportation.
\section * {ACNOWLEDGMENTS}
This project was supported by the National Natural Science
Foundation of China (Project NO. 10174066).

\end{document}